\documentclass[11pt]{article}
\usepackage{graphicx}
\begin{document}
\begin{center}

{\Large{\bf Magnetic field decay and period evolution of the source RX
J0720.4-3125\\}}
{\it by S.B.  Popov and D.Yu.  Konenkov.  }

Izvestiya VUZov "Radiofizika" V.41 P.28-35 (1998)\\
(English translation: Radiophysics and quantum electronics)\\

Presented at the summer school Volga-97 (1-11 June 1997)\\

\vskip 0.5cm

\end{center}

\begin{center}
{\Large{ Abstract}}
\end{center}

We studied possible evolution of rotational period and  magnetic
field of the X-ray source RX J0720.4-3125 assuming this source to be an
isolated neutron star accreting from interstellar medium.
Magnetic field of the
source is estimated to be $10^6 - 10^9$ Gs (most probably
$\approx 2\cdot 10^8$ Gs), and it is difficult to
explain the observable rotational period  8.38 s without invoking
hypothesis of the magnetic field decay. We used the model of ohmic
decay of crustal magnetic field. The estimates of the accretion
rate ($10^{-14} - 10^{-16} M_\odot/{\rm yr}$), velocity of the source
relative to the interstellar medium ($10 - 50 $ km/s), the neutron star
age ($2\cdot 10^9 - 10^{10}$ yrs) are obtained.\\

\vskip 1cm

\noindent
(remark added in proof: As we found out in July 1997 when this paper was
already submitted, common results were recieved independently by John C.L.
Wang // Astrophys. J., 1997, V.486. P.L119)

\newpage

\section{Introduction}

Recently isolated neutron stars (INSs), which are not observed as
radiopulsars, became of great interest both for observers and
theorists. The idea of observations of such sources was proposed
more than 25 years ago [6], and in 1991 Treves and Colpi [8] suggested,
that INSs, accreting from  interstellar medium (ISM), can be
observed in great number 
in UV and X-rays by $ROSAT$. Here we present the results
of the work on the source RX J0720.4-3125 observed by Haberl et al. [12].
In [12] it was proposed,
that RX J0720.4-3125 is an accreting INS with the rotational period
8.38 s.

There are four possible stages of an INS in  low density
plasma:  Ejector (E), Propeller (P), Accretor (A) and Georotator (G).
The stage is determined by relations between specific radii:
$R_l$-- light cylinder radius, $ R_{st}$-- stopping radius,
$R_G=(\frac {2\, G\, M}{v_\infty ^2})$-- the radius of the gravitational
capture and $R_{co}=(\frac {G\, M}{\omega ^2})^{1/3}$-- the corotation
radius.

As a result we have two critical periods: $P_E$ and $P_A$,
separating different stages. If $p<P_E$, we have an Ejector
(i.e. a pulsar), if $P_E<p<P_A$, the NS is on the Propeller stage,
and if  $p>P_A$ and $R_{st}<R_G$ -- we have an accreting NS.
It is possible, that $p>P_A$, but $R_{st}>R_G$. In this case
a geo-like magnitosphere is formed, and we call it Georotator.

For describtion of period evolution it is useful to use a
gravimagnetic parameter, y, (it is described in [4]).

On the figure 1 three examples of evolutionary tracks of an INS on p-y-
diagram are presented. All of them are ended at the Accretor
stage.

\noindent
 I. $E\longrightarrow P\longrightarrow A$-- evolution of an INS in the
ISM with the constant density without magnetic field decay.

\noindent
 II. $E\longrightarrow P\longrightarrow A \longrightarrow P \longrightarrow A$--
evolution of an INS, passing through a giant molecular cloud
without  magnetic field decay.

\noindent
 III. Evolution with magnetic field decay.

We are especially interested in the third case.


\section{Analitical estimates of the X-ray pulsar's parameters}

In this part we show mainly estimates which were not included
into the article [3].

Lets consider the situation, when the Alfven radius, $R_A$, and
the corotation radius, $R_{co}$, are equal. This equation
gives us the accretion period, $P_{A}$:

 \begin{equation}
   P_A\approx 6\cdot10^2\mu_{30}^{6/7}\rho_{-24}^{-3/7}
         v_{\infty_6}^{9/7}\left(\frac {M}{M_{\odot}}\right)^{-11/7}\, s
 \end{equation}

\noindent
here $\mu_{30}$ -- magnetic momentum in units
$10^{30}{\rm Gs \cdot cm}^2$, $\rho_{-24}$ -- density of the ISM
in units $10^{-24}{\rm g/cm}^3$, $v_{\infty_6}$ -- velocity
of the INS relative to the ISM in units $10^6 {\rm cm/s}$.

During accretion $p >P_A$. If these two periods are equal:

$$
  \mu_{30}^{6/7}=\frac{8.38}{6\cdot 10^2} \rho_{-24}^{3/7}
                 v_{\infty_6}^{-9/7}\left(\frac M{ M_{\odot}}\right)^{11/7}
$$

So, $\mu_{30}=0.007 \rho_{-24}^{1/2}v_{\infty_6}^{-3/2}
\left(\frac M{ M_{\odot}}\right)^{11/6}$
for $P_A=p=8.38 \,\, s$ (the period of RX J0720.4-3125).
For the NS's radius $R=10\,{\rm km}$ we have $B<7\cdot10^{9}$ Gs.

 If we use the hypothesis of the acceleration of an INS from the
turbulizated ISM, we have an equation, which gives an
estimate of the magnetic field of the INS based on the
equilibrium period, $P_{eq}$, [3, 5]:

\begin{equation}
P_{eq}=2355 k_t^{1/3}\mu_{30}^{2/3}I_{45}^{1/3}\rho_{-24}^{-2/3}
v_{\infty_6}^{13/3}v_{t_6}^{-2/3}
\left(\frac M{M_{\odot}}\right)^{-8/3}\,\, {\rm s}
\end{equation}

Here
$I_{45}$ -- the momentum of inertia in units $10^{45}{\rm g\cdot cm}^2$,
$v_{t_6}$ -- the turbulent velocity in units $10^6{\rm cm/s}.$

    So we have:
$$
  \mu_{30}=\left(\frac{8.38}{2355}\right)^{3/2} k_t^{-1/2}I_{45}^{-1/2}
           \rho_{-24}v_{\infty_6}^{-13/2}v_{t_6}\left(\frac M{M_{\odot}}\right)^4
$$

\noindent
    And finaly $B\approx 2.1\cdot 10^8$ Gs.

 In [3] we showed, that the magnetic field can't be less than
$10^5-10^6\, Gs$,
because in the opposite case we can't observe pulsations
of the X-ray flux.

\begin{figure}
\includegraphics[width=12cm]{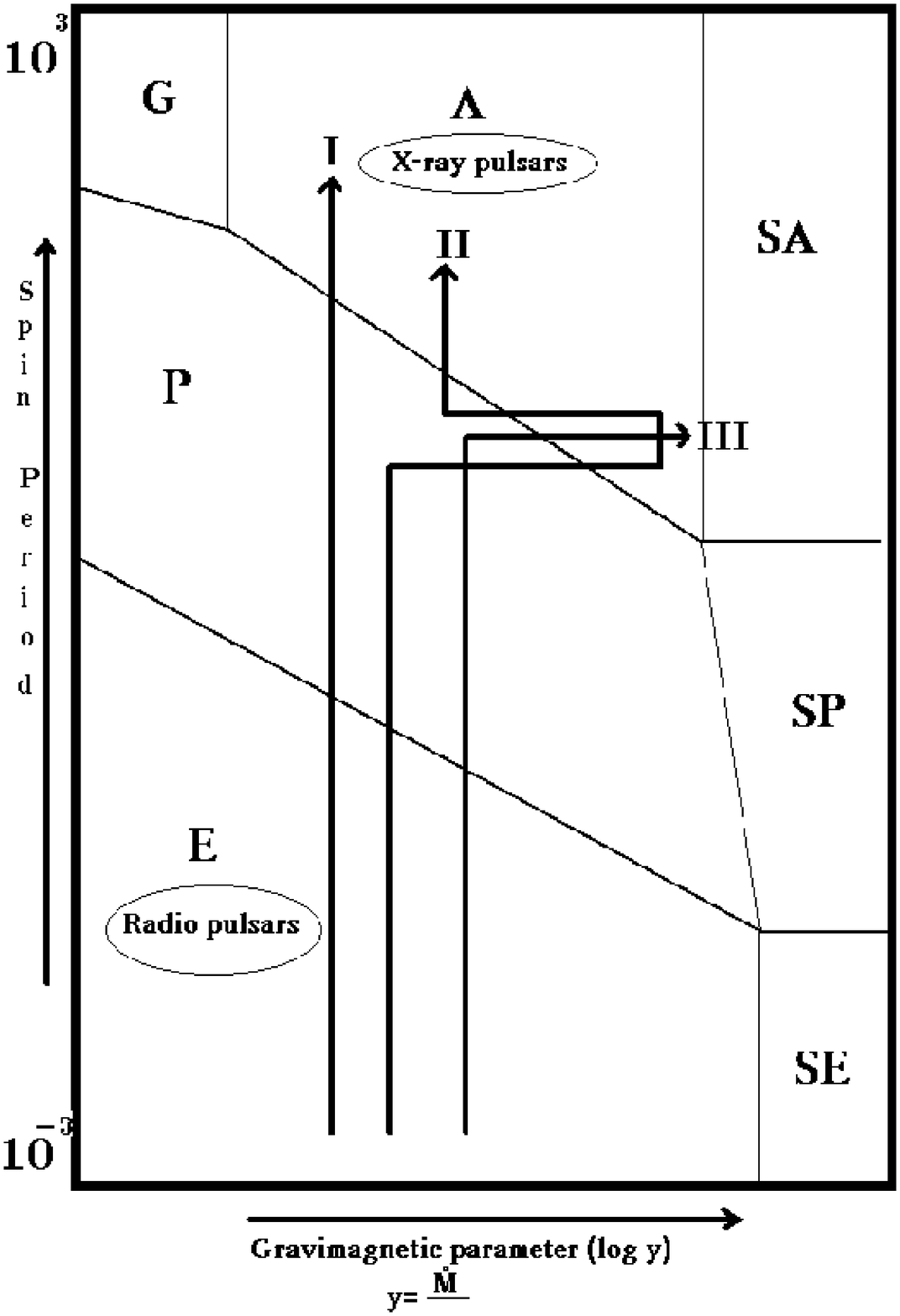}
\caption{ P-y diagram}
\end{figure}

\begin{figure}
\includegraphics[width=12cm]{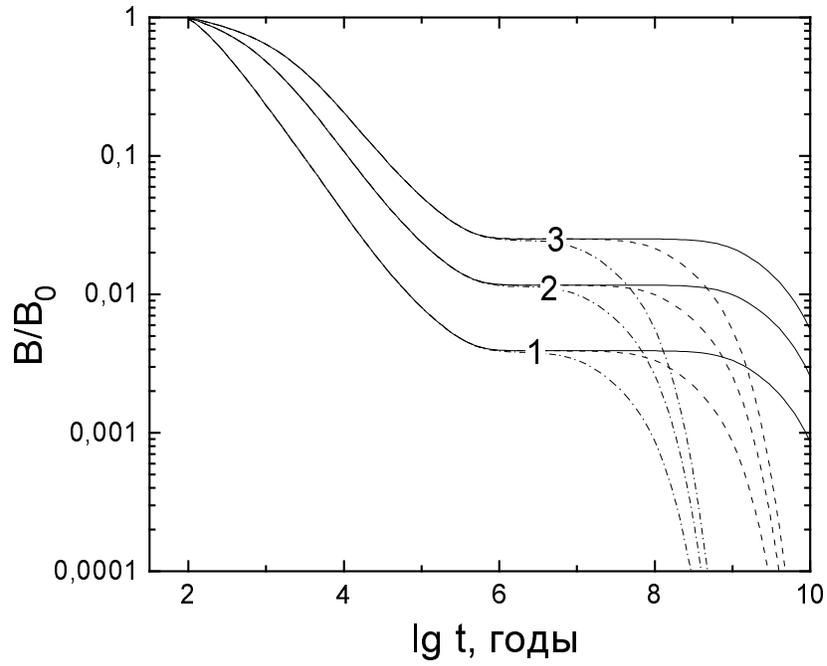}
\caption{ 
Changes of the surface magnetic field of an isolated neutron star 
with time for the model of standard cooling. Curves 1, 2, and 3 correspond 
to the initial depths of the current layer, $10^{11}$, $10^{12}$, and
$10^{13}\,$ g cm$^{-3}$, respectively. The solid curves correspond to 
Q = 0.001; the dashed curves, to Q = 0.01; the dot-dashed curves, to Q = 0.1. 
}
\end{figure}

\begin{figure}
\includegraphics[width=12cm]{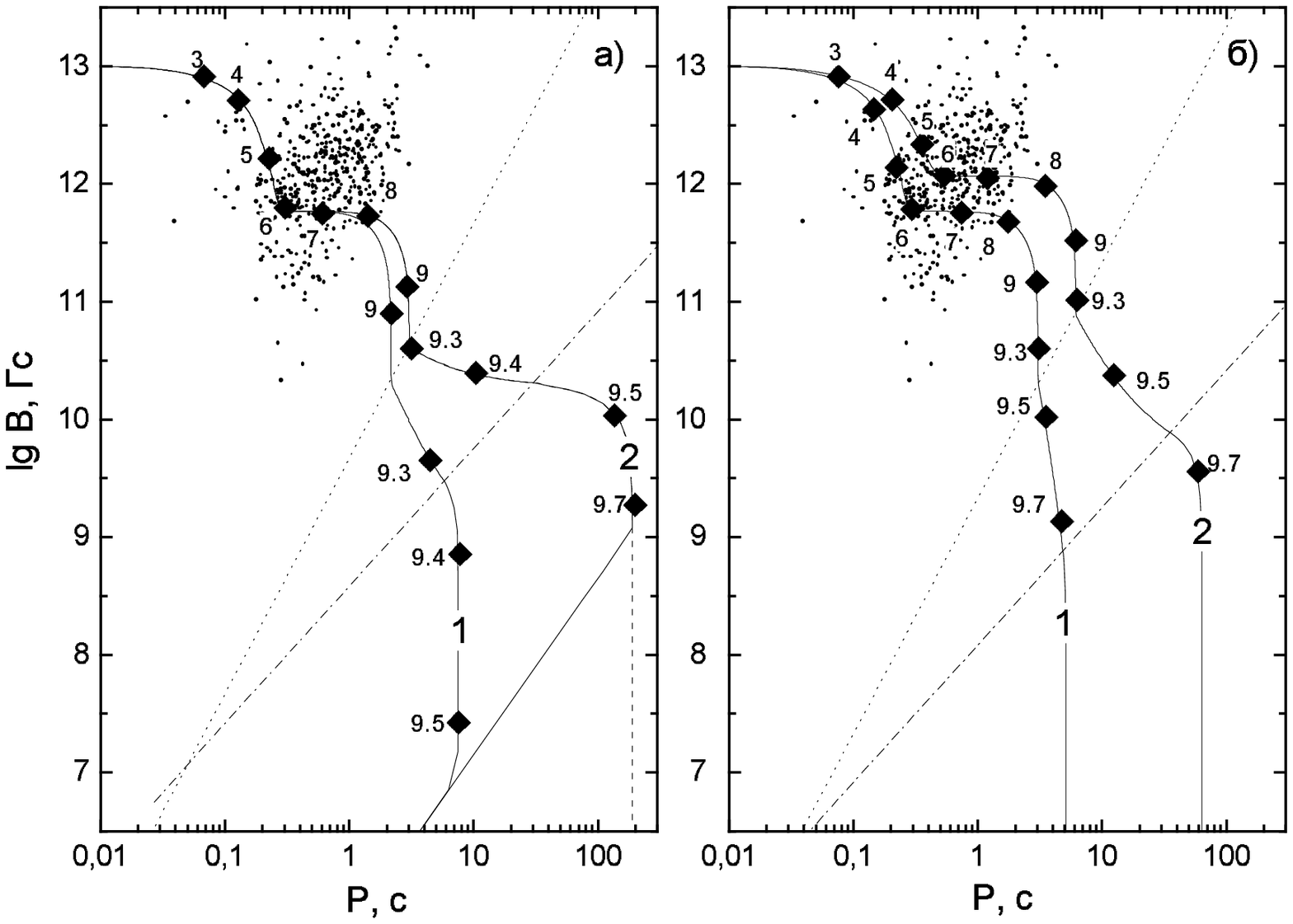}
\caption{ 
 The evolutionary tracks of the neutron star for the accretion rates
$\dot M = 
10^{-15} M_{\odot}\,$yr$^{-1}$ (a) and $\dot M = 10^{-16} M_{\odot}\,$
yr$^{-1}$ (b). The model 
parameters are described in the text. The dashed lines correspond to $p = 
P_E$; the dot-dashed lines, to $p = P_A$. The dashed line in Fig. 2a 
shows for the second track the neutron star evolution with no acceleration in 
the turbulized intestellar medium. The numbers near the marks in tracks 
denote the logarithm of the neutron star age in years. The observed radio 
pulsars are indicated by dots. 
}
\end{figure}

Lets suppose, that the INS was born with a period about 0.01 -- 0.02 s
(the exact value is not very important, because the only necessary
condition is $P_{initial} <<$  8.38 s, and initially
the INS was on the Ejector stage). So, the star had to decelerate
till 8.38 s. Lets estimate the time of this deceleration.

The time of deceleration is determined by the {\it final}, not by the
initial period!

$$
  P_E\approx 10 (k_t)^{1/4} \mu_{30}^{1/2} \rho_{-24}^{-1/4} v_{\infty_6}^{-1/2} s
$$

$$
  \frac{dI\omega}{dt}=-k_t\frac{\mu^2}{R_l^3},
  \, R_l=c/\omega, \, \omega=2\pi/p
$$

$$
  \frac1{p^2}I\frac{\Delta p}{\Delta t}=k_t\frac{\mu^2(2\pi)^2}{c^3p^3}
$$

    So for $\Delta p=p$:

$$
  \Delta t=\frac{Ip^2c^3}{k_t\mu^2(2\pi)^2}=
$$
$$
  =3\cdot 10^7 P_E^2k_t^{-1}\mu_{30}^{-2}I_{45} yrs=
$$
$$
  =3\cdot 10^9 k_t^{-1/2}\mu_{30}^{-1}\rho_{-24}^{-1/2}v_{\infty_6}^{-1}I_{45} yrs
$$

    For $\mu_{30}< 0.01 \,\,\, \Delta t > 3\cdot 10^{11} yrs >> t_{Hubble}$.
I.e. the star couldn't initially have low magnetic field.

As far as the NS is on the Accretor stage, its period should be
much greater than the ejection period, $P_E$:

$$
  p>P_{Propeller}>P_E
$$

$P_E\approx 10 s$ for standart parameters. It means, that
if the star is on the accretor stage with p=8.38 s, the field
is much less than the standart value: $B << 10^{12}$ Gs.

Lets try to estimate the characteristic time of acceleration
and deceleration of such a NS:

$$
  t_{su}=t_{sd}=\frac{I\omega}{\dot M v_t R_G}=
$$

$$
  =20 yrs \cdot I_{45}v_{\infty_6}^5\rho_{-24}^{-1}
   \left(\frac p{10^5 s}\right)^{-1}\left(\frac {v_t}{10^6 cm/s}\right)^{-1}
$$
where $t_{su}, t_{sd}$ - characteristic times of the period changes.

    For p=10 s $t_{su}=2\cdot 10^5 $ yrs. If  $v_t$ is less than
$10^6 cm/s$, than $t_{su}$ is greater. So, $\dot p \approx
p/t_{su} < \approx \frac{8.38}{2\cdot 10^5
\cdot 3 \cdot 10^7} \approx 10^{-12} s/s$


%


\section{Calculations of the magnetic field decay}

 So, we can say, that the magnetic field of the source dissipated,
and the characteristic time of the dissipation , $t_d$, was short enough:
$t_d < t_E$ ($t_E$-- Ejector's time). Because of that we have rapidly
rotating INS on the Accretor stage.

The detailes of calculations one can find in [3].

We calculated the magneto-rotational
evolution of the star with the mass
$M=1.4 M_\odot$ for accretion rates $10^{-15}M_\odot {\rm /yr}$ and
$10^{-16}M_\odot {\rm /yr}$ passing through the ISM with the density
$\rho=10^{-24}{\rm g/cm^3}$.  These rates correspond to
velocities $\approx 20$ and $\approx 40$ km/s.

The magnetic field decay was studied by different authors
(see [10]).For conductivity we used formulae from [13] and [2].
Initially the magnetic field was localized in the surface layer.
Low accretion rates don't have much influence on the
magnetic field decay [1, 9].

On the figure 2 we show the decrease of the surface magnetic field
for different parameters. We used the model of the NS structure from
[11].

On figure 3 the evolutionary tracks for the accretion rates
$10^{-15} M_\odot {\rm /yr}$ (fig. 3a) and
$10^{-16} M_\odot {\rm /yr}$ (fig. 3b) are shown.


\section{Conclusions}

Observations of the accreting INSs can be an important test
for theories of the magnetic field decay. In that case
we have a very "pure" example of the decay, because a lot of
effects of the interaction of the magnetic field, the surface
of the NS and surrounding plasma are not important.

Observations of the source
RX J0720.4-3125 [12] showed the existence of the INS with the rotation period
which can be easily explained with the assumption of the
magnetic field decay. It can be important for estimates
of the total number of observable INS and for their appearence
as periodic X-ray sources [7].


We want to thank dr.F. Haberl for the information about the source,
drs. M.E. Prokhorov, V.A. Urpin, D.G. Yakovlev and V.M. Lipunov for helpful
discussion, and participantes of the summer school
 "The many faces of neutron stars", where the idea of this article was born.

The work of D.K. was supported by RFFI 96-02-16905a,
the work of S.P. -- by RFFI 95-02-06053 , INTAS 93-3364,
ISSEP a96-1896.

\section{References}

\noindent
1. Zdunik J.L.,
Haensel P., Paczy\'nski B., Miralda-Escude J.// Astrophys. J.
1992. V.384. P.129.

\noindent
2. Itoh N., Hayashi H., Kohyama Y.
// Astrophys. J. 1993. V.418. P.405.

\noindent
3. Konenkov D.Yu, Popov S.B.//
 Astronomy Letters. 1997. V.23, P. 200 (astro-ph/9707318)

\noindent
4. Lipunov V.M.// {\it Astrophysics of neutron stars},
Springer, 1992.

\noindent
5. Lipunov V.M., Popov S.B.//
 Astron. Zhur.. 1995. V.72, P. 711. (see also astro-ph/9609185)

\noindent
6. Ostriker J.P., Rees M. J., Silk J. //
 Astrophys. J. Letters 1970. V.6, P.179

\noindent
7. Popov S.B.//Astron. Circ 1994. No.1556. P.1.

\noindent
8. Treves A., Colpi M.//
 Astron. and Astrophys., 1991, V. 241, P. 107.

\noindent
9. Uprin V., Geppert
U., Konenkov D.// Astron. and Astrophys. 1996. V.307. P.807.

\noindent
10. Urpin V.A., Muslimov A.G.// Astron. Zhur. 1992. V.69.
P.1028.

\noindent
11. Friedman B., Pandharipande V.R.//
Nucl. Phys. 1981. A361. P.502.

\noindent
12. Haberl F., Pietsch W., Motch C., Buckley D.A.H.//
IAU Circ. 1996. No. 6445.

\noindent
13. Yakovlev D.G., Urpin V.A.// Astron. Zhur. 1980. V.24. P.303.

\end{document}